\documentclass[12pt]{article}
\usepackage[margin=1in]{geometry}

\usepackage{subfigure}
\usepackage{amsmath,amssymb}

\usepackage{amsthm} 
\allowdisplaybreaks
\usepackage{color}
\usepackage{footnote}
\usepackage{algorithm}
\usepackage{algpseudocode}
\usepackage{algorithmicx}
\usepackage{multirow} 
\usepackage{booktabs}
\usepackage{graphicx}
\usepackage{amssymb}
\usepackage{amsbsy}
\usepackage{array}
\usepackage{longtable}
\usepackage{epstopdf}
\usepackage{pbox}
\usepackage{url}
\usepackage{breqn}
\usepackage{mathrsfs}
\usepackage{multicol}
\usepackage{supertabular}
\usepackage{enumerate}
\usepackage{bbm}
\usepackage{multirow}
\usepackage{hyperref}
\usepackage{cite}
\usepackage{xfrac}
\usepackage{textgreek}
\usepackage[justification=centering]{caption}
\usepackage{mathtools}

\usepackage{todonotes}
\usepackage{dsfont}
\usepackage{bm}
\usepackage{makecell}
\usepackage[ruled,vlined,linesnumbered,algo2e]{algorithm2e}
\algnewcommand\algorithmicforeach{\textbf{for each}}
\algdef{S}[FOR]{ForEach}[1]{\algorithmicforeach\ #1\ \algorithmicdo}

\begin{document}
\title{\LARGE \bf A Multi-Agent Rollout Approach for Highway Bottleneck Decongestion in Mixed Autonomy}

\author{Lu Liu, Maonan Wang, Man-On Pun and Xi Xiong
\thanks{This work was supported in part by NSFC Project 72371172 and Fundamental Research Funds for the Central Universities.}
\thanks{L. Liu and X. Xiong are with the Key Laboratory of Road and Traffic Engineering, Ministry of Education, Tongji University, Shanghai, China.  M. Wang and Man-On Pun are with the School of Science and Engineering, the Chinese University of Hong Kong, Shenzhen, China (Emails: luliu0720@tongji.edu.cn, maonanwang@link.cuhk.edu.cn, simonpun@cuhk.edu.cn, xi\_xiong@tongji.edu.cn,)}%
}
\newcommand*{\QEDA}{\hfill\ensuremath{\blacksquare}}%
\date{}
\maketitle

\begin{abstract}
The integration of autonomous vehicles (AVs) into the existing transportation infrastructure offers a promising solution to alleviate congestion and enhance mobility. This research explores a novel approach to traffic optimization by employing a multi-agent rollout approach within a mixed autonomy environment. The study concentrates on coordinating the speed of human-driven vehicles by longitudinally controlling AVs, aiming to dynamically optimize traffic flow and alleviate congestion at highway bottlenecks in real-time. We model the problem as a decentralized partially observable Markov decision process (Dec-POMDP) and propose an improved multi-agent rollout algorithm. By employing agent-by-agent policy iterations, our approach implicitly considers cooperation among multiple agents and seamlessly adapts to complex scenarios where the number of agents dynamically varies. Validated in a real-world network with varying AV penetration rates and traffic flow, the simulations demonstrate that the multi-agent rollout algorithm significantly enhances performance, reducing average travel time on bottleneck segments by 9.42\% with a 10\% AV penetration rate. 
\end{abstract}

{\bf Keywords}:
Autonomous vehicles, traffic flow control, multi-agent rollout, reinforcement learning.

\section{INTRODUCTION}


The potential of autonomous vehicles (AVs) in enhancing road capacity, reducing stop-and-go traffic, and optimizing overall traffic flow has positioned them as key players in intelligent transportation systems \cite{Chang1997AnalysisOC}. With the acceleration of urbanization, traffic congestion has emerged as a critical challenge, imposing substantial economic and productivity costs. Congestion often occurs near bottlenecks, which are identified as primary contributors to the \textit{Capacity Drop} \cite{Shaikh2022ARO, Chung2007RelationBT}, due to their relatively low capacity and susceptibility to traffic overloading. Thus, by controlling the number of vehicles prior to bottlenecks, traffic congestion can be alleviated. Although various control methods have been proposed, ramp control is the most widely applied solution \cite{su151612608}. However, traffic flow on congested highway segments often originates from multiple entrances, ramp control models are often tailored to local segments and may not be effective. The potential of AVs in intelligent traffic management offers new opportunities for addressing traffic congestion \cite{adler2016optimal}.

\begin{figure*}[!h]
    \centering
    \includegraphics[width=1.0\linewidth]{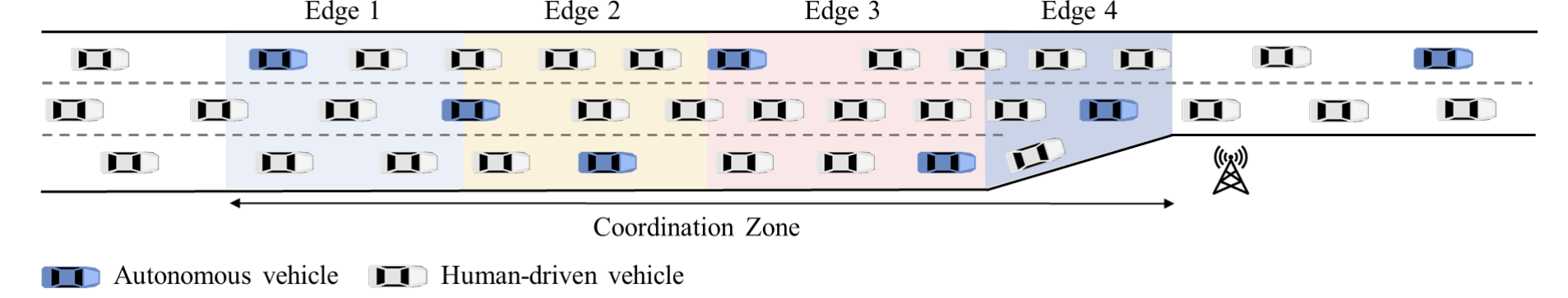}
    \caption{Illustration of congestion where three lanes merge into two, with congestion originating downstream (right) and propagating upstream (left).}
    \label{Figure 2}
\end{figure*}

In this paper, we tackle the challenge of mitigating traffic congestion in dynamic environments by controlling the speed of AVs to coordinate the timing of upstream vehicles approaching bottlenecks. Considering the scenario with mixed autonomy traffic flow depicted in Fig.~\ref{Figure 2}, where a bottleneck occurs as three lanes merge into two, blue vehicles represent AVs, and white vehicles represent human-driven vehicles (HDVs). In uncontrolled areas, AVs emulate the driving behavior of HDVs. Upon entering the controlled coordination zone, AVs utilize onboard cameras and satellite navigation systems (e.g., GPS) to gather self-information, including their position, speed, and distance to the bottleneck exit. Furthermore, to enhance situational awareness, the controlled coordination zone is subdivided into multiple edges, allowing AVs to access the vehicle count on each lane of every edge through connectivity established with roadside cameras via vehicular communication systems. Based on these data, strategic decision-making collaboration among AVs ensues, determining which vehicle should accelerate to traverse the bottleneck promptly and which should decelerate and wait to pass, thereby reducing a reduction in the average travel time of vehicles on the segment.

Existing research that exploits the speed control of upstream AVs has demonstrated potential in mitigating congestion \cite{Nie2021VariableSL}. The methods of inter-vehicle communication (IVC) and road-vehicle communication (RVC) are widely used to manage AVs \cite{xiong2021optimizing}. However, this relies on existing infrastructure and cannot guarantee the success of communication. Li et al. \cite{LiEnhanced2020} proposed a model-based predictive control framework to reduce the travel time of vehicles in a network. The model-based control methods require rigorous and explicit models of traffic flow dynamics to ensure optimal control. Most of the literature uses model-free approaches to model the problem of mitigating traffic congestion using AVs. Maske et al. \cite{Maske2019} introduced a distributed approach, with each AV independently responsible for its control area. While relying solely on distributed control often overlooks the intricacies of multi-vehicle cooperation. Consequently, multi-agent RL algorithms \cite{Rashid2018RTS} have been proposed and implemented for the coordinated control of AVs \cite{Wang2024AMR}. Although these algorithms allow AVs to interact, they struggle to scale with the increasing number of agents, often leading to computational bottlenecks and suboptimal performance in real-world scenarios. 

To address this gap, our study formulates the control tasks of multiple AVs as a decentralized partially observable Markov decision process (Dec-POMDP) and presents a novel multi-agent rollout approach. This approach reframes the multi-agent decision-making problem into a sequential decision-making paradigm using the agent-by-agent policy iteration (A2PI) \cite{Bertsekas2021}, and updates each AV's policy via a Proximal Policy Optimization (PPO) algorithm \cite{Schulman2017ProximalPO}. By enabling the actions of preceding AVs to inform the state inputs of subsequent AVs, this method fosters an inherently adaptive cooperative interaction model. Crucially, our approach is designed to flexibly accommodate dynamic fluctuations in the number of AVs, ensuring robust performance in evolving traffic conditions. The main contributions of this article can be summarized as follows:
\begin{itemize}
    \item We model the multi-agent problem as a Dec-POMDP to capture the intricate interactions among agents and provide a unified framework for decision-making in multi-agent systems.
    \item A multi-agent rollout algorithm is developed, utilizing A2PI to convert the multi-agent problem into a sequential decision-making problem. It is demonstrated that this iterative approach consistently enhances the policy, effectively promoting collaboration among agents and adapting smoothly to changes in the number of agents. 
    \item The approach has been tested in mixed autonomy traffic flow on a real-world network. Experimental results indicate a significant 9.42\% improvement in the average travel time of vehicles on bottleneck segments with a 10\% penetration rate of AVs.
\end{itemize}

The remainder of this paper is organized in the following manner. Section~\ref{02_modeling} introduces the model for the scenario depicted in Fig.~\ref{Figure 2}. Section~\ref{03_multiagent} details the enhanced multi-agent rollout approach. Section~\ref{04_experiments} analyzes the performance of the proposed approach through simulations based on real traffic scenarios. Finally, Section~\ref{05_conclusions} summarizes the main findings and outlines several directions for future work. 
\section{Modeling and Formulation} \label{02_modeling}

In this section, the methodology for optimizing traffic flow by incorporating AVs is presented within a Dec-POMDP framework. The objective is to minimize travel costs for all vehicles in the network. Consider a highway segment with a bottleneck, as illustrated in Fig.~\ref{Figure 2}, where both AVs and HDVs converge, leading to a reduction in the highway's capacity. In the Dec-POMDP model, each AV is limited to its own sensor data and the information disseminated by the infrastructure, thus possessing a partial view of the overall traffic situation. Prior to the bottleneck, AVs undertake control actions, which in this study are confined to longitudinal speed adjustments to simplify computational demands.

The traffic management challenge for a fleet of $N$ AVs approaching a bottleneck is structured within the Dec-POMDP paradigm. This paradigm is characterized by the tuple $\left(\mathcal{S}, \{ \mathcal{A}_i \}_{i \in \mathcal{N}}, \{ \mathcal{O}_i \}_{i \in \mathcal{N}}, \mathcal{T}, \mathcal{R}, \gamma, \tau \right)$, where $\mathcal{N}$ represents the number of the agents, $\mathcal{S}$ represents the set of global states of the environment, inaccessible to the AVs, $\{\mathcal{A}_i\}$ is the set of actions available to each AV $i$, and $\mathcal{A} = \times_i \mathcal{A}_i$ defines the joint action space. Similarly, $\{\mathcal{O}_i\}$ denotes the set of observations for each AV, with $\mathcal{O} = \times_i \mathcal{O}_i$ representing the joint observation space. The state transition probability function $\mathcal{T}: \mathcal{S} \times \mathcal{A} \times \mathcal{S} \rightarrow [0, 1]$ governs the likelihood of transitioning between states given the actions taken by the AVs. The reward function $\mathcal{R}: \mathcal{S} \times \mathcal{A} \rightarrow \mathbb{R}$ measures the immediate utility of actions, while $\gamma \in [0,1]$ is the discount factor, and $\tau$ specifies the planning horizon. The goal is to optimize the cumulative long-term reward by coordinating AV acceleration and deceleration actions. 

At each discrete time step $t$, AV $i$ receives an observation $\bm{o}_i^t\in \mathcal{O}_i$, and all AVs execute a joint action $\bm{a}^t = \{a^t_1, \cdots, a^t_N\} \in \mathcal{A}$, resulting in a state transition according to $\mathcal{T}(\bm{s}^t,\bm{a}^t,\bm{s}^{t+1})$. Subsequently, each AV $i$ receives a collective reward $r^t$ is issued based on $\mathcal{R}$. The total reward is the discounted expected sum over an infinite horizon, with the optimal policy derived from the Bellman equation:
\begin{equation}
    J^\ast(\bm{s}^t)=\max_{\bm{a}^t \in \mathcal{A}} \left[r^t(\bm{s}^t,\bm{a}^t) + \gamma \, \phi^\ast(\bm{s}^{t+1})\right],
\end{equation}
where $\phi^\ast(\bm{s}^{t+1})=\sum\limits_{\bm{s}^{t+1} \in \mathcal{S}}\mathcal{T}(\bm{s}^t,\bm{a}^t,\bm{s}^{t+1}) J^\ast(\bm{s}^{t+1})$ encapsulates the expected return from the next state, thus connecting the present decision-making to future rewards.

For effective vehicle control before the bottleneck, the specific Dec-POMDP representation is as follows. At any given time step $t$, each AV agent, indexed by $i$ where $i = 1,2, \ldots, N$, acquires kinematic data including its velocity $v_i^t$ and the distance $d_i^t$ to the bottleneck's exit. To incorporate global traffic information, road segment data is encapsulated by the distribution of vehicles across edges and lanes, with $n_{e,l}^t$ denoting the number of vehicles on edge $e \in \{1,2, \ldots, E\}$ and lane $l \in \{1,2, \ldots, L\}$, where $L$ and $E$ represent the total number of lanes and edges. The observation state for vehicle $i$ is thus formalized as:

\begin{equation}
    \bm{o}_i^t = \left[\begin{array}{cc}
    \underbrace{v_i^t, d_i^t}_{\text{AV information}} & \underbrace{n_{1,1}^t, n_{1,2}^t, \ldots, n_{E,L}^t}_{\text{Road segment information}}
    \end{array}\right].
\end{equation}

Agent $i$ selects actions $a_i^t$ in the form of acceleration commands $\eta_i^t$, which are bounded within the operational limits $\left[\eta_{\min}, \eta_{\max}\right]$. The reward function $r^t$ is designed to reflect the average speed $v_b$ of the vehicles within the control zone, thereby aligning individual agent objectives with the overarching goal of enhancing traffic flow and minimizing congestion. Importantly, the reward incorporates a time penalty to discourage AVs from reducing speed or halting in unregulated areas solely to accumulate rewards. 
\section{Multi-agent Rollout Approach for Traffic Flow Optimization} \label{03_multiagent}

In this section, a modified multi-agent rollout method based on A2PI is delineated for optimizing traffic flow within environments of mixed autonomy. Initially, the concept of A2PI is elucidated, and it is substantiated that iterative refinement of agent strategies enhances their performance. Subsequently, the multi-agent rollout algorithm combining A2PI with PPO \cite{Schulman2017ProximalPO} is explained, which is used to fine-tune traffic flow optimization.

\subsection{Agent-by-agent policy iteration}

In the proposed A2PI approach, a joint policy vector $\bm{\pi} = (\pi_1, \pi_2, ..., \pi_N)$ governs the behavior of all AVs. Each AV, indexed by $i$ where $i \in \{1, 2, 3, ..., N\}$, is initially equipped with a policy $\pi_{i}^{0}$, which is a stochastic strategy generated at random, with the superscript $k=0$ denoting the iteration index. As the iteration process unfolds, the policy $\pi_{i}$ of a singular AV is selected for refinement, while the policies of its counterparts are maintained unchanged. Such a sequential training methodology is designed to alleviate the instability that is frequently encountered in multi-agent reinforcement learning paradigms.

To facilitate effective cooperation among multiple AVs, the policy $\pi_{i}$ of an AV is optimized using the cumulative reward of all AVs, which promotes the enhancement of the policy through iterative updates. This approach enables an AV to predict and adapt to the collective behavior of the group, which is crucial for achieving collaborative objectives.
\begin{figure}[H]
\vspace{-0.5em}
    \centering
    \includegraphics[width=0.7\linewidth]{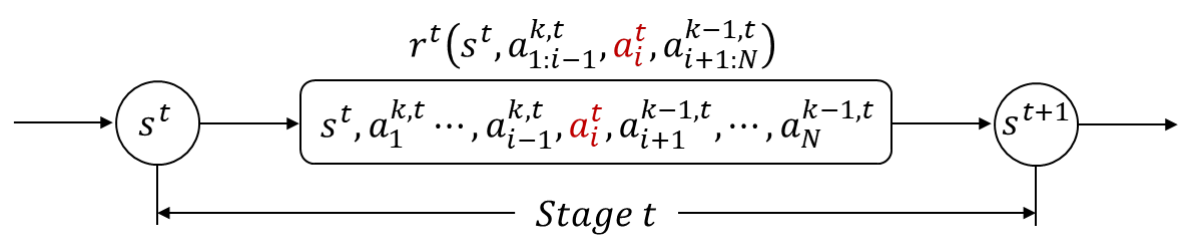}
    \caption{Sequential policy update in multi-agent decentralized control.}
    \label{Figure 3.}
\end{figure}
\vspace{-0.5em}
Fig.~\ref{Figure 3.} illustrates the update process for a single AV's policy. It is noted that AVs undergo policy iteration in a sequential manner. For the $k$-th iteration of AV $i$ at time $t$, the policies of AVs $1$ to $i-1$ have undergone $k$ iterations, whereas the policies of AVs $i$ to $N$ have completed $k-1$ iterations. The joint policy configuration at this stage can be represented as follows, where $\bm{\pi}^{k(i-1)}$ indicates the policy for AVs $1$ through $i-1$ after the $k$-th iteration. Once all AV policies have undergone $K$ iterations, the configuration will be denoted as $\bm{\pi}^{k(i)}$, and for brevity, as $\bm{\pi}^{k}$:

\begin{equation}
    \bm{\pi}^{k(i-1)} = \{\pi_{1}^{k}, \ldots, \pi_{i-1}^{k}, \pi_{i}^{k-1}, \ldots, \pi_{N}^{k-1}\}.
\end{equation}

Based on their respective observations $\bm{o}^{t}_{j}$ for $j \in \{1,2,\cdots,N\}$, AV $i$ chooses the action $a_{i}^{t}$ to maximize the reward, and other AVs select actions according to their strategies: $a_{j}^{k,t}$ for $j \in \{1,2, \ldots, i-1\}$ and $a_{j}^{k-1,t}$ for $j \in \{i+1, \ldots, N\}$. This selection results in a collective reward $r^t(s^{t}, a_{1:i-1}^{k,t}, a_{i}^{t}, a_{i+1:N}^{k-1,t})$, where $a_{1:i-1}^{k,t}$ is the action tuple $(a_{1}^{k,t}, \ldots, a_{i-1}^{k,t})$ and $a_{i+1:N}^{k-1,t}$ is $(a_{i+1}^{k-1,t}, \ldots, a_{N}^{k-1,t})$. During this process, only the policy model $\pi_{i}^{k}$ is updated, with the parameters of other strategies remaining unchanged. Upon completion of the update, $\pi_{i}^{k}$ is obtained. Subsequently, AV $i+1$ proceeds to update its policy, and this process continues until the policies of all agents have been updated, culminating in the optimization of the overall system's performance. The effectiveness of this approach is substantiated in the following proposition.

\textbf{Proposition 1.} For the states $\bm{s}^t \in \mathcal{S}$, the joint policy $\bm{\pi}^{k+1}$ obtained in iteration $k+1$ is not inferior to the policy $\bm{\pi}^{k}$ from iteration $k$, i.e.,
\begin{equation}
    J_{\bm{\pi}^{k}}(\bm{s}^t) \leq J_{\bm{\pi}^{k+1}}(\bm{s}^t), \quad \forall s^t \in \mathcal{S}.
\end{equation}

\textbf{Proof:} To demonstrate the validity of Proposition 1, consider a simplified multi-agent system comprising two agents (i.e., $N = 2$). The expected reward for state $\bm{s}^t$ under the joint policy $\bm{\pi}^{k}$ is given by:

\begin{align}
    J_{\bm{\pi}^{k}}(\bm{s}^t) &= \mathbb{E} \left[ r^t(\bm{s}^t, a^{k,t}_1, a^{k,t}_2) + \gamma \phi_{(\pi^{k}_1, \pi^{k}_2)}(\bm{s}^{t+1}) \right] \nonumber \\
    &\leq \max_{a_1^{t} \in \mathcal{A}_1} \mathbb{E} \left[ r^t(\bm{s}^t, a_1^{t}, a^{k,t}_2) + \gamma \phi_{(\pi^{k}_1, \pi^{k}_2)}(\bm{s}^{t+1}) \right] \nonumber \\
    &= \mathbb{E} \left[ r^t(\bm{s}^t, a_1^{k+1,t}, a^{k,t}_2) + \gamma \phi_{(\pi^{k+1}_1, \pi^{k}_2)}(\bm{s}^{t+1}) \right] \nonumber \\
    &\leq \max_{a_2^{t} \in \mathcal{A}_2} \mathbb{E} \left[ r^t(\bm{s}^t, a_1^{k+1,t}, a_2^{t}) + \gamma \phi_{(\pi^{k+1}_1, \pi^{k}_2)}(\bm{s}^{t+1}) \right] \nonumber \\
    &= \mathbb{E} \left[ r^t(\bm{s}^t, a_1^{k+1,t}, a_2^{k+1,t}) + \gamma \phi_{(\pi^{k+1}_1, \pi^{k+1}_2)}(\bm{s}^{t+1}) \right] \nonumber \\
    &= J_{\bm{\pi}^{k+1}}(\bm{s}^t).
\end{align}

In the above derivation, the expected reward under the joint policy $\bm{\pi}^{k}$ is first established. Subsequently, the policy for the first agent is improved, as indicated by the maximization operation, ensuring that the expected reward following the policy enhancement for the first agent is at least as favorable as the previous one. The new policy for the first agent is then adopted. The subsequent steps mirror the process for the second agent, culminating in the validation of Proposition 1.

It is thereby demonstrated that in a two-agent system, the expected reward obtained when agents employ the optimal policies is at least equivalent to the expected reward under the original policies. This substantiates the efficacy of the proposed approach in augmenting the average velocity at traffic bottlenecks. The proof can be extended analogously for any positive integer number of agents $N$.

\subsection{Multi-agent rollout algorithm}

Building upon the foundation of the A2PI approach, we integrate the PPO in the multi-agent rollout algorithm to refine the policy of each AV within the mixed traffic flow. PPO stands out for its stable and efficient policy updates, which are critical in a policy gradient approach that aims to maximize expected returns. It achieves this by limiting the magnitude of policy modifications, a feature that is well-suited to the multi-agent rollout method based on A2PI. This compatibility is due to PPO's ability to facilitate gradual and consistent policy enhancement for each agent, thereby ensuring that the collective system progresses towards optimal policy convergence without introducing instability to the learning trajectory.

Incorporating the PPO algorithm with the A2PI framework involves a meticulous synchronization of the agent-wise policy updates with the PPO's objective. For agent $i$ at the $k$-th iteration, the local environmental information is obtained through observations $\bm{o}_i$, and the policy $\pi_{i}^{k}$ is refined by maximizing the PPO objective function, which is defined as:  

\begin{align}
    L(\pi_{i}^{k}) = \hat{\mathbb{E}}_t \Big[ \min \Big( r_t(\pi_{i}^{k}) \hat{A}_t,\nonumber \text{clip}\big(r_t(\pi_{i}^{k}), 1-\epsilon, 1+\epsilon\big) \hat{A}_t \Big)\Big],
\end{align}

where $r_t(\pi_{i}^{k})$ is the probability ratio $r_t(\pi_{i}^{k}) = \frac{\pi_i^k(\bm{a}_t|\bm{o}^t_i)}{\pi_{i}^{k-1}(\bm{a}_t|\bm{o}^t_i)}$, signifying the likelihood of selecting action $\bm{a}_t$ under the new policy relative to the old policy. The advantage function estimate at time $t$ is represented by $\hat{A}_t$, and $\epsilon$ is a hyperparameter controlling the clipping range. The empirical average over a finite batch of samples is indicated by $\hat{\mathbb{E}}_t$. This clipped objective allows for a controlled policy update that maintains the benefits of policy iteration while mitigating the risks of policy performance collapse. 

The PPO algorithm is further adapted to the multi-agent rollout approach by updating the policies of the agents sequentially, as depicted in the A2PI process. Each policy $\pi_i^k$ is improved by collecting a set of trajectories under the current policy configuration $\bm{\pi}^k$, computing advantage estimates, and then applying the PPO update rule. The sequential update is crucial for maintaining the stability of the learning process in the multi-agent environment, as it allows each agent's policy to adapt gradually to any new behaviors emerging from the collective policy changes of all agents. The pseudocode for the multi-agent rollout algorithm is shown in Algorithm~\ref{Multi-Agent Rollout}.

\begin{algorithm}[ht]
\caption{Multi-Agent Rollout for Traffic Flow Optimization}
\label{Multi-Agent Rollout}
\DontPrintSemicolon
\KwIn{Initial policies $\pi_i^0$ for all AVs $i \in \{1, 2, ..., N\}$.}
\KwOut{Converged policies $\pi_i^*$ for all AVs $i$.}
\BlankLine
Initialize policies $\pi_i^0$ for all AVs $i \in \{1, 2, ..., N\}$.\;
\For{$k=0,1,2,...$ \KwTo convergence}{
    \For{each agent $i \in \{1, 2, ..., N\}$}{
        Collect set of trajectories $\mathcal{D}_i$ under $\bm{\pi}^k$.\;
        Compute advantage estimates $\hat{A}_t$ for each step in $\mathcal{D}_i$.\;
        \For{epoch $= 1,2,...,M$}{
            Update $\pi_i^k$ by maximizing $L(\pi_i^k)$ using $\mathcal{D}_i$ and $\hat{A}_t$.\;
        }
        Update $\pi_i^k$ to $\pi_i^{k+1}$.\;
    }
}
\end{algorithm}

In this algorithm, the collection of trajectories $\mathcal{D}_{i}$ is performed by sampling actions from the current policy $\pi_{i}^{k}$ and interacting with the environment. The advantage function estimates $\hat{A}_{t}$ are computed to quantify the benefit of the selected actions over the baseline, which guides the policy update. The policy update itself is performed for $M$ epochs to ensure sufficient optimization of the PPO objective within the trust region defined by the clipping mechanism. This iterative process is repeated until the policies converge to an optimal solution, thereby optimizing the traffic flow and reducing congestion within the mixed autonomy environment.

The multi-agent rollout algorithm presented in this study embodies a decentralized control strategy. Each AV selects its optimal action through a process of distributed learning, which involves implicit cooperation among all agents. Our method's robustness is highlighted by its scalability to various numbers of agents; it adapts to changes in agent populations without the need for re-engineering the decision-making framework. Moreover, the policy iteration for each agent is conducted in a manner where the other agents' policies remain stationary, promoting a stable learning backdrop that is essential for the convergence of individual training processes. This stability is a significant advantage, addressing the frequent convergence difficulties faced in multi-agent learning algorithms due to environmental complexity. 
\section{Numerical Results} \label{04_experiments}

\subsection{Experiment setting}
This study conducted experiments utilizing the Simulation of Urban Mobility platform(SUMO). RL-based AV optimization was facilitated by the Adam optimizer. The configuration parameters were set as follows: a learning rate of $1 \times 10^{-4}$, a discount factor $\gamma$ of 0.9, and a clipping hyperparameter $\epsilon$ of 0.2. The evaluation metric was the average travel time (Avg.TT), where lower values indicate better performance.

\begin{figure}[ht]
    \centering
    \subfigure[The map of the selected segments on the Shenhai Highway]{
        \includegraphics[width=0.65\linewidth]{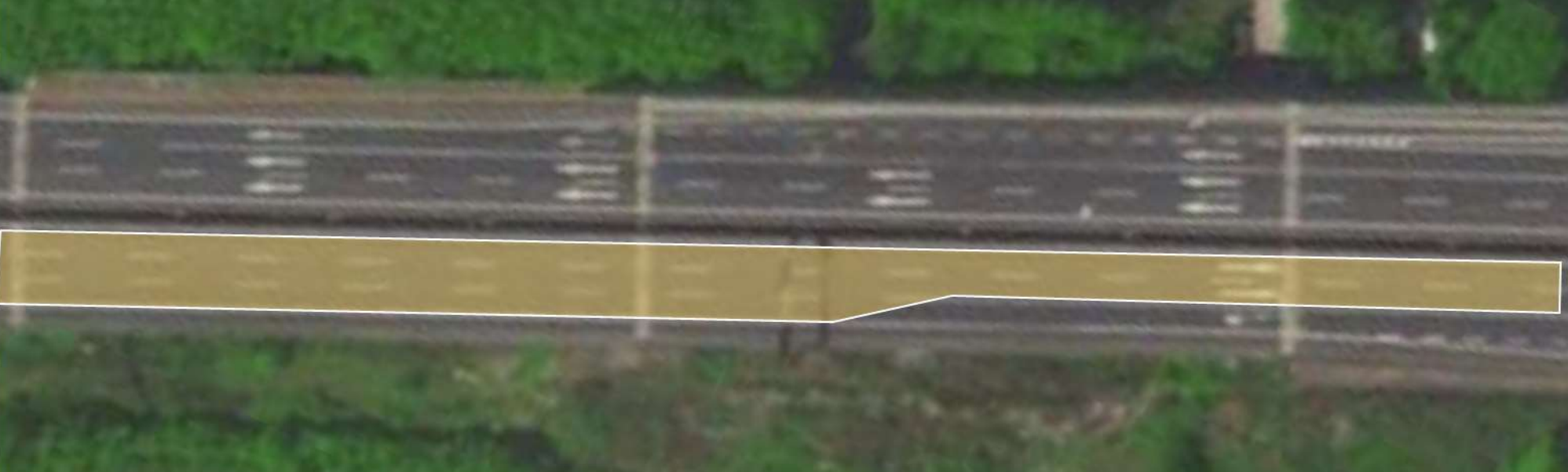}
        \label{fig:shenhai_map}
    }
    \subfigure[Simulation of the Shenhai Highway bottleneck segment]{
        \includegraphics[width=0.65\linewidth]{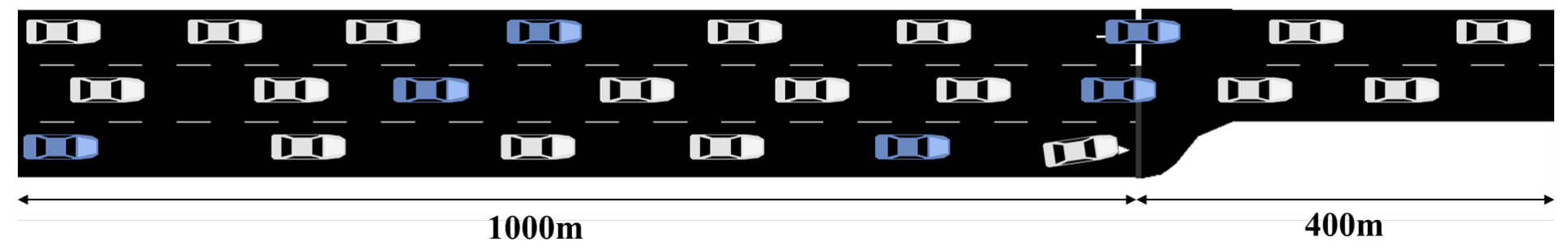}
        \label{fig:sumo_simulation}
    }
    \caption{Traffic flow optimization at a bottleneck on the Shenhai Highway.}
    \label{Figure 4.}
\end{figure}

A real-world traffic scenario on the Shenhai Highway in Shanghai, characterized by a segment where three lanes merge into two, was selected to assess the proposed method's efficacy in alleviating congestion at the bottleneck. The chosen coordination zone, extending over 1,000~$m$ and depicted in Fig.~\ref{Figure 4.}, illustrates both the map of the selected areas on the Shenhai Highway and the simulation of the bottleneck segment in SUMO. Two traffic scenarios were set up, a congestion scenario with a simulation lasting 700~$s$, where the traffic flow is 2400~$veh/hour$ including 6 AVs, and another scenario with a simulation lasting 7200~$s$ where traffic flow varies randomly from 0 to 4000~$veh/hour$ and 10\% of the vehicles are AVs. Maximum speeds were established at 10~$m/s$ for HDVs and 12~$m/s$ for AVs. The action space of an AV was defined as a set of accelerations from $\eta_{\min}$=$-5$~$m/s^s $ to $\eta_{\max}$=$2.5$~$m/s^s $. Vehicles were initially released randomly from the start of a designated roadway, and allowed to travel 400~$m$ freely. By controlling AVs' speed to enable vehicles in specific lanes to decelerate and allow others to overtake, the overall waiting time is reduced.

\subsection{Compared methods}

The performance of the algorithm proposed in this paper is evaluated against three benchmark methods:

\begin{itemize}
    \item Without Control: In this method, the vehicles determine their speed within the SUMO simulation environment by considering the distance to the vehicle ahead, the desired headway, and so on \cite{Treiber2000CongestedTS}. In this paper, without control is used as a baseline methodology.
    
    \item MATD3: While the Twin Delayed DDPG (TD3) algorithm is designed for single-agent tasks, it has been adapted to multi-agent systems \cite{Vinitsky2020OptimizingMA} by providing a separate TD3 instance for each agent.
    
    \item MAPPO: The design of the Multi-Agent Proximal Policy Optimization (MAPPO) algorithm 
 \cite{Yu2021TheSE}, which utilizes a global value function to evaluate the collective state of all agents, illustrates its positive role in coordinating cooperation among multiple agents.
\end{itemize}

\subsection{Performance comparison}

Table~\ref{table 3} presents the performance of various methods across two traffic scenarios. In the first scenario, characterized by congestion with a traffic flow of 2400 $veh/hour$ including 6 AVs, the multi-agent rollout algorithm reduces the average travel time by 45.13\% after two iterations compared to scenarios without control. Additionally, the final total reward obtained by the multi-agent rollout algorithm is higher than those achieved by MATD3 and MAPPO. In the second scenario, where traffic flow varies randomly from 0 to 4000 $veh/hour$, including a 10\% AV penetration, the algorithm demonstrates a 9.42\% improvement in average travel time compared to uncontrolled conditions. Here too, the final reward of our method surpasses those of MATD3 and MAPPO. 

\begin{table}[ht]
\centering
\setlength{\belowcaptionskip}{0.2cm}
\caption{Performance comparisons of different algorithms}
\label{table 3}
\setlength{\tabcolsep}{1.2mm}{
\begin{tabular}{ccccc}
\Xhline{1.0pt}
Scenario &Algorithm &Reward &Avg.TT (s) & Perf (\%)\\
\hline
\multirow{4}*{T=700s} & Without Control &-20537 & 563.46 & --  \\
~ &MATD3 &-4248 & 347.34 & 38.36  \\
~ &MAPPO &-3405 & 311.16 & 44.78\\  
~ &MARollout & \textbf{-3273} & \textbf{309.17} & 45.13\\
\hline
\multirow{4}*{T=7200s} &Without Control &-41540 & 168.72 & -- \\
~ &MATD3 &-26691 & 157.95 &6.38 \\
~ &MAPPO &-22315 & 154.67 & 8.33\\  
~ &MARollout & \textbf{-20826}  & \textbf{152.82} & 9.42 \\
\Xhline{1.0pt}
\end{tabular}}
\parbox{\textwidth}{
\vspace{2mm}
\hspace*{8pt}
\begin{tabular}{@{}p{\textwidth}@{}}
{$\quad \quad \quad \quad \quad \quad^a$} \textit{T} denotes simulation time;\\
{$\quad \quad \quad \quad \quad \quad^b$}\textit{Perf} denotes performance improvement;\\
{$\quad \quad \quad \quad \quad \quad^c$} \textit{MARollout} denotes multi-agent rollout algorithm
\end{tabular}
}
\end{table}

\begin{figure}[ht]
    \centering
    \includegraphics[width=0.5\linewidth]{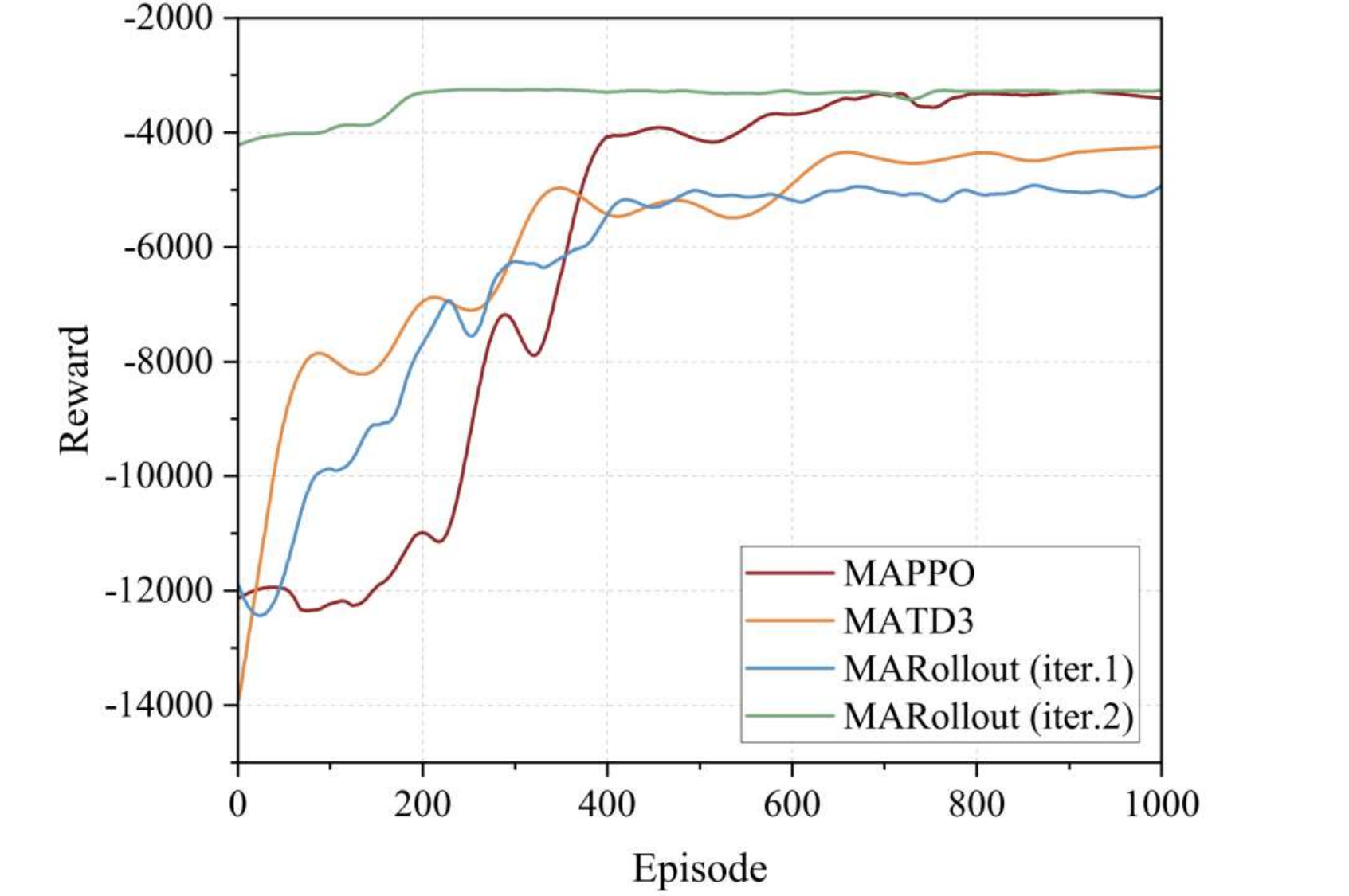}
    \caption{The performance curves of the algorithms during training in the first scenario. MARollout (iter.1) and MARollout (iter.2) represent the results of 1-step and 2-step iteration, using the multi-agent rollout approach, respectively.}
    \label{Figure 5}
\end{figure}

\begin{figure}[ht]
    \centering
    \includegraphics[width=0.6\linewidth]{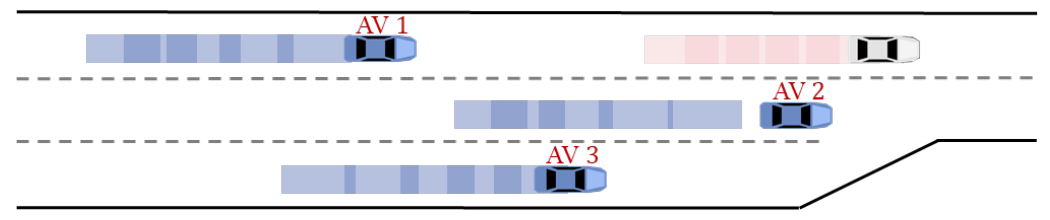}
    \caption{An illustration of the cooperative actions of AVs near bottlenecks.}
    \label{Figure 6}
\end{figure}

Fig.~~\ref{Figure 5} illustrates the reward curves of different algorithms in the first traffic scenario. It is observed that after two iterations, the MARollout achieves the highest reward. The evolution of rewards over the two iterations of the multi-agent rollout algorithm indicates the development of cooperation among agents. Initially, the total reward for all agents begins at a relatively low baseline within a stochastic environment. By the second iteration, there is a significant increase in the initial reward, exceeding the final reward of the first iteration. This improvement suggests that the agents developed effective policies during the first iteration to coordinate their actions. Fig.~\ref{Figure 6} illustrates this cooperative. We selected four vehicles near the bottleneck at a movement and depicted their action variations in trajectory form over the first six time steps. Among the three AVs, AV~$2$ was passing through the bottleneck unaffected by other AVs, with its acceleration gradually increasing. Through observation, AV~$1$ identified only one vehicle ahead in its lane and has no competition with AV~$2$, so executing acceleration actions. Due to AV~$3$ observing the presence of AV~$2$ at the bottleneck, it decelerated in advance to avoid waiting.

\vspace{-1.0em}

Although the performance of the multi-agent rollout algorithm was comparable to MAPPO, the MAPPO assumes the number of agents is fixed at the start of training, which makes it powerless in dynamic scenarios with agents entering and exiting the system frequently. Moreover, as the number of agents increases and policies vary, the environment is unstable, which makes it difficult for MAPPO to converge. In contrast, utilizing a policy iteration approach, the multi-agent rollout algorithm adeptly addresses these challenges.

\subsection{Sensitivity analysis}

\begin{figure}[ht]
    \centering
    \subfigure[Average travel time with the inflow]{\includegraphics[width=0.48\linewidth]{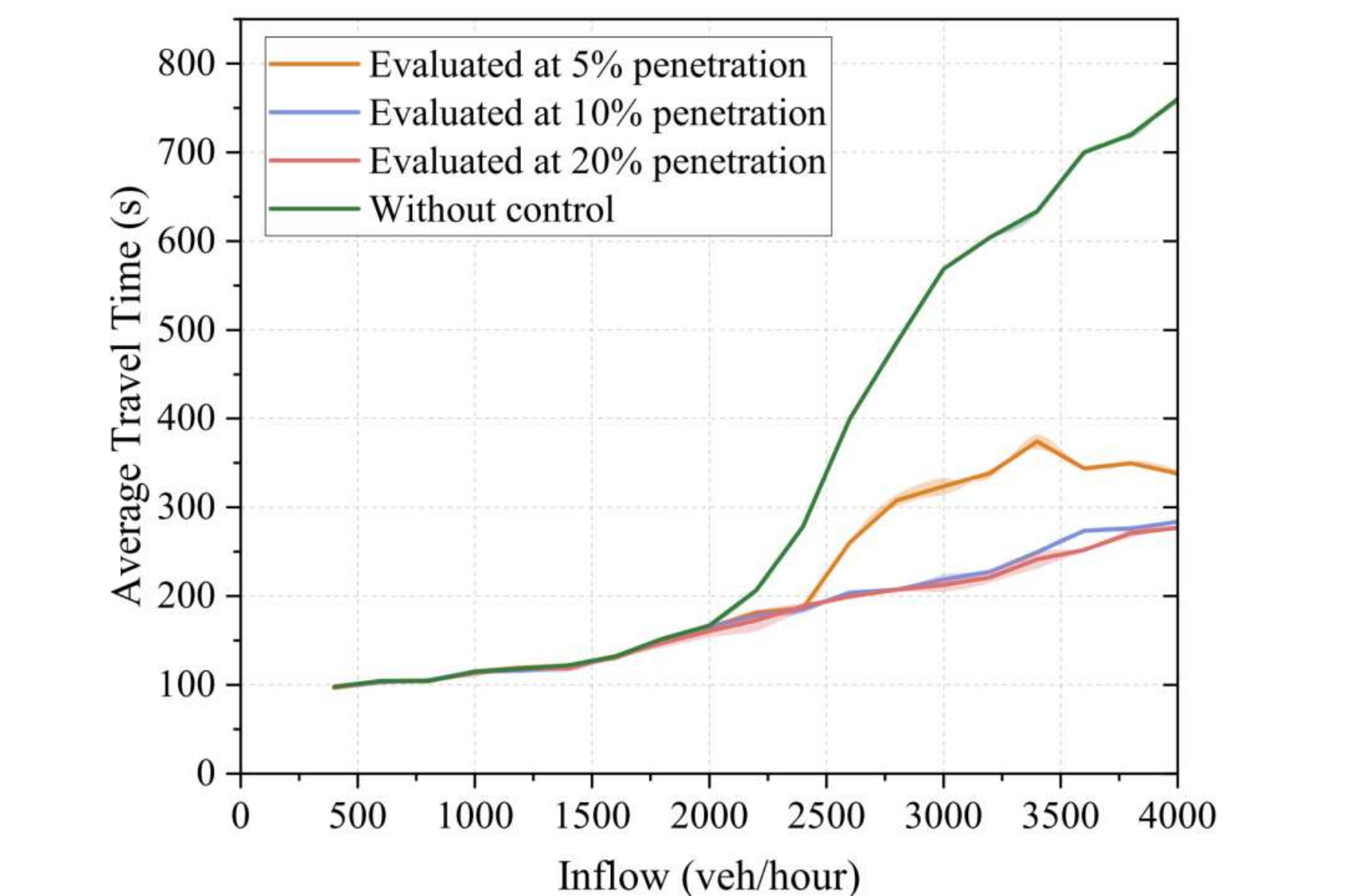}}
    \subfigure[Average travel time with the penetration rate]{\includegraphics[width=0.48\linewidth]{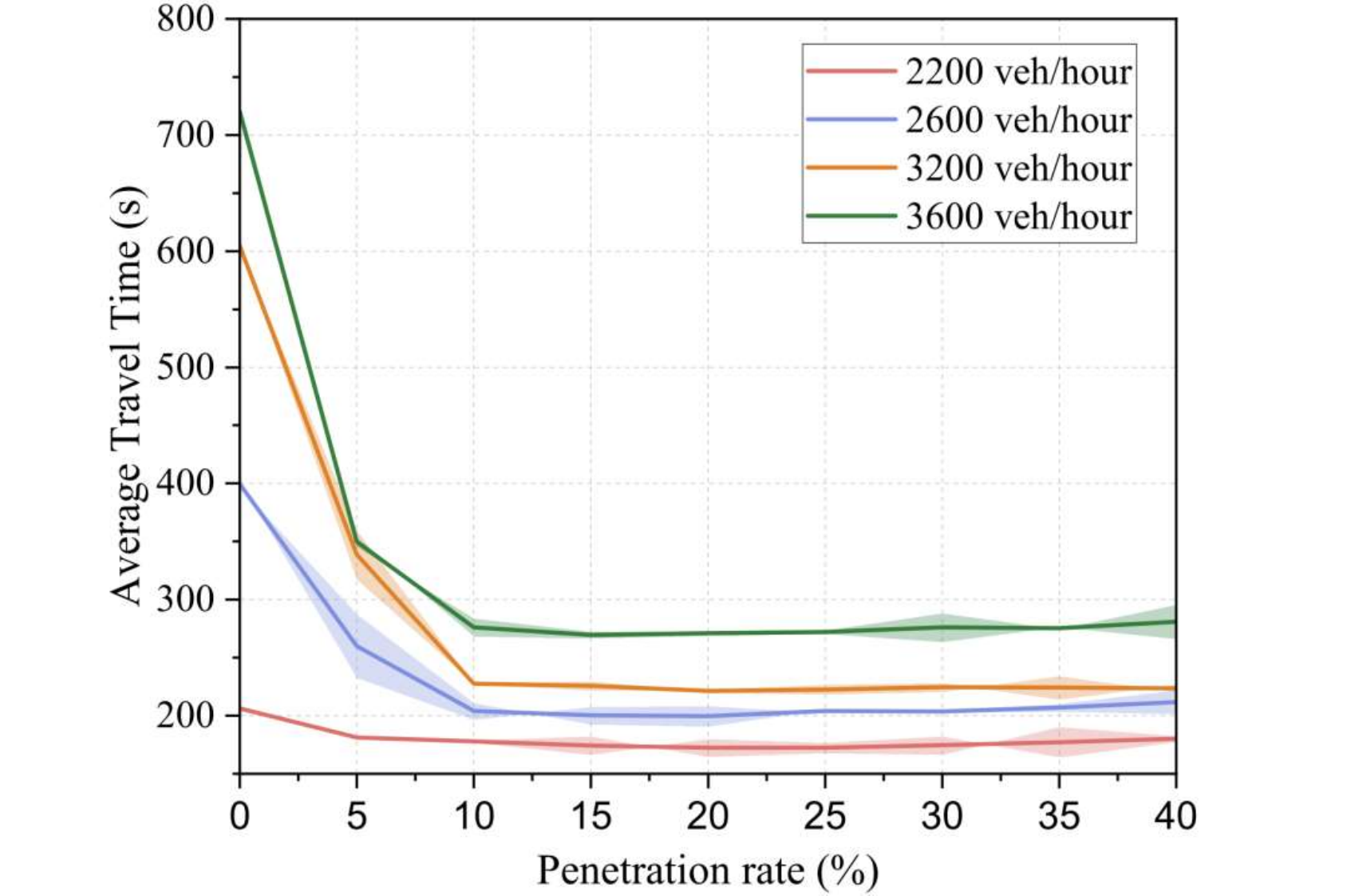}}
    \caption{Sensitivity analysis of average travel time to changes in autonomous vehicles penetration rate and traffic inflow.}
    \label{Figure 7}
\end{figure}

This subsection examines the sensitivity of average travel time to key parameters, including traffic inflow and AV penetration rate. We have conducted a sensitivity analysis of inflow at 5\%, 10\%, and 20\% penetration rates of AVs, respectively. As shown in Fig. \ref{Figure 7}(a), the average travel time increases with inflow for all penetrations. Nevertheless, the presence of AVs significantly reduces the average travel time, especially when the inflow exceeds 2,000~$veh/hour$, underscoring the AVs' capability to preemptively traffic conditions at the bottleneck and slow down to avoid congestion.

Subsequently, the influence of AV penetration rate on average travel time is shown in Fig.~\ref{Figure 7}(b). The results illustrate that a penetration rate range of 0\% to 40\% yields significant average travel time reductions across all traffic flow. Notably, beyond the 10\% penetration rate, the travel time begins to a stabilized value, indicating that the penetration benefits of AVs may be nearing saturation. 
\section{CONCLUSIONS}
\label{05_conclusions}

In this paper, we demonstrate the effectiveness of coordinated control of autonomous vehicles in alleviating traffic congestion at bottlenecks within mixed autonomy traffic flow. By modeling the problem as a Dec-POMDP and utilizing the multi-agent rollout approach based on A2PI, the task of multi-agent control is transformed into sequential decision-making distributed control. With the AV penetration rate of 10\%, the average travel time at bottlenecks is reduced by 9.42\%. This method not only enhances cooperation among agents but also addresses the challenge of adapting to changes in agent numbers, a common limitation of multi-agent algorithms.

This work can be extended in several directions. Firstly, incorporating vehicles' lane-changing behavior could reflect real-world traffic flow more accurately. Secondly, applying this method to larger-scale networks would allow for a deeper investigation into how multiple agents impact traffic flow within complex networks. Thirdly, optimizing the policy update process by training multiple agents in parallel would have faster optimization.
\bibliographystyle{plain}
\bibliography{optimization}
\end{document}